%% file: Main.tex
\documentclass[conference]{IEEEtran}
\IEEEoverridecommandlockouts

\usepackage{cite}
\usepackage{amsmath,amssymb,amsfonts}
\usepackage{algorithmic}
\usepackage{graphicx}
\usepackage{textcomp}
\usepackage{tabularx}
\usepackage{soul}

\usepackage[table]{xcolor}  
\definecolor{lightblue}{RGB}{128,255,255}  
\definecolor{headercolor}{gray}{0.8}  

\definecolor{lightcyan}{RGB}{128, 230, 255} 

\definecolor{lightpink}{RGB}{255, 220, 220} 

\newcommand{\highlightcyan}[1]{%
    \tikz[baseline=(X.base)]{
        \node[fill=cyan!30,rectangle,rounded corners,inner sep=2pt] (X) {#1};
    }%
}

\newcommand{\highlightpink}[1]{%
    \tikz[baseline=(X.base)]{
        \node[fill=pink!70,rectangle,rounded corners,inner sep=2pt] (X) {#1};
    }%
}

\usepackage{xargs} 
\usepackage[colorinlistoftodos,prependcaption,textsize=small]{todonotes}
\def\BibTeX{{\rm B\kern-.05em{\sc i\kern-.025em b}\kern-.08em
    T\kern-.1667em\lower.7ex\hbox{E}\kern-.125emX}}

\usepackage{tikz}               
\usepackage{amsmath}     
\usepackage{wrapfig}
\usepackage{sidecap}
\usepackage{caption}
\usepackage{subcaption}

\usepackage{soul}               
\usepackage{booktabs}
\usepackage{comment}            
\usepackage{enumitem}           
\usepackage{amsthm}                 
\usepackage{thmtools}               
\usepackage{algorithmic}            
\usepackage[ruled,vlined,linesnumbered]{algorithm2e} 
\usepackage{url} 
\usepackage{multicol}               
\usepackage[euler]{textgreek}       
\usepackage{multirow}           
\usepackage{framed}

\usepackage{adjustbox}
\usepackage{multirow}
\usepackage{tabularx}
\usepackage{lipsum}

\usepackage{arydshln}
\usepackage{soul}
\usepackage{tikz}
\usetikzlibrary{calc}
\usetikzlibrary{decorations.pathmorphing}
\usepackage{xpatch}

\makeatletter

\newcommand{\defhighlighter}[3][]{%
  \tikzset{every highlighter/.style={color=#2, fill opacity=#3, #1}}%
}

\defhighlighter{yellow}{.5}

\newcommand{\highlight@DoHighlight}{
  \fill [ decoration = {random steps, amplitude=1pt, segment length=15pt}
        , outer sep = -15pt, inner sep = 0pt, decorate
        , every highlighter, this highlighter ]
        ($(begin highlight)+(0,8pt)$) rectangle ($(end highlight)+(0,-3pt)$) ;
}

\newcommand{\highlight@BeginHighlight}{
  \coordinate (begin highlight) at (0,0) ;
}

\newcommand{\highlight@EndHighlight}{
  \coordinate (end highlight) at (0,0) ;
}

\newdimen\highlight@previous
\newdimen\highlight@current

\DeclareRobustCommand*\highlight[1][]{%
  \tikzset{this highlighter/.style={#1}}%
  \SOUL@setup
  \def\SOUL@preamble{%
    \begin{tikzpicture}[overlay, remember picture]
      \highlight@BeginHighlight
      \highlight@EndHighlight
    \end{tikzpicture}%
  }%
  \def\SOUL@postamble{%
    \begin{tikzpicture}[overlay, remember picture]
      \highlight@EndHighlight
      \highlight@DoHighlight
    \end{tikzpicture}%
  }%
  \def\SOUL@everyhyphen{%
    \discretionary{%
      \SOUL@setkern\SOUL@hyphkern
      \SOUL@sethyphenchar
      \tikz[overlay, remember picture] \highlight@EndHighlight ;%
    }{%
    }{%
      \SOUL@setkern\SOUL@charkern
    }%
  }%
  \def\SOUL@everyexhyphen##1{%
    \SOUL@setkern\SOUL@hyphkern
    \hbox{##1}%
    \discretionary{%
      \tikz[overlay, remember picture] \highlight@EndHighlight ;%
    }{%
    }{%
      \SOUL@setkern\SOUL@charkern
    }%
  }%
  \def\SOUL@everysyllable{%
    \begin{tikzpicture}[overlay, remember picture]
      \path let \p0 = (begin highlight), \p1 = (0,0) in \pgfextra
        \global\highlight@previous=\y0
        \global\highlight@current =\y1
      \endpgfextra (0,0) ;
      \ifdim\highlight@current < \highlight@previous
        \highlight@DoHighlight
        \highlight@BeginHighlight
      \fi
    \end{tikzpicture}%
    \the\SOUL@syllable
    \tikz[overlay, remember picture] \highlight@EndHighlight ;%
  }%
  \SOUL@
}
\makeatother

\DeclareRobustCommand{\IEEEauthorrefmark}[1]{\smash{\textsuperscript{\footnotesize #1}}}

\begin{document}
\title{Lessons from the Use of Natural Language Inference (NLI) in  Requirements Engineering Tasks
}

\author{
    \IEEEauthorblockN{Mohamad Fazelnia\IEEEauthorrefmark{1}, Viktoria Koscinski\IEEEauthorrefmark{2}, Spencer Herzog\IEEEauthorrefmark{2}, Mehdi Mirakhorli\IEEEauthorrefmark{1}}
    \IEEEauthorblockA{\IEEEauthorrefmark{1}Department of Computer Science, University of Hawaii at Manoa\\
    Email: mfazel@hawaii.edu, mehdi23@hawaii.edu}
    \IEEEauthorblockA{\IEEEauthorrefmark{2}Global Cybersecurity Institute, Rochester Institute of Technology\\
    Email: vk2635@rit.edu, seh2132@rit.edu} 
}

\maketitle

\begin{abstract}

We investigate the use of Natural Language Inference (NLI) in automating requirements engineering tasks. In particular, we focus on three tasks: requirements classification, identification of requirements specification defects, and detection of conflicts in stakeholders' requirements. While previous research has demonstrated significant benefit in using NLI as a universal method for a broad spectrum of natural language processing tasks, these advantages have not been investigated within the context of software requirements engineering. Therefore, we design experiments to evaluate the use of NLI in requirements analysis.  We compare the performance of NLI with a spectrum of approaches, including prompt-based models, conventional transfer learning, Large Language Models (LLMs)-powered chatbot models, and probabilistic models. Through experiments conducted under various learning settings including conventional learning and zero-shot, we demonstrate conclusively that our NLI method surpasses classical NLP methods as well as other LLMs-based and chatbot models in the analysis of requirements specifications. 
Additionally, we share lessons learned characterizing the learning settings that make NLI a suitable approach for automating requirements engineering tasks.

\end{abstract}



\begin{tikzpicture}[remember picture,overlay]
  \node[anchor=north, yshift=-1cm] at (current page.north) {\normalsize 2024 IEEE 32nd International Requirements Engineering Conference};
\end{tikzpicture}
\vspace*{-20pt} 

\begin{tikzpicture}[remember picture,overlay]
  \node[anchor=south, align=center, font=\tiny] at ($(current page.south)+(0,1cm)$) {
    \textcopyright \ \textcopyright \ 20xx IEEE. Personal use of this material is permitted.
    Permission from IEEE must be obtained for all other uses, in any current or future media,
    including reprinting/republishing this \\ material for advertising or promotional purposes,
    creating new collective works, for resale or redistribution to servers or lists, or reuse of any
    copyrighted component of this work in other works.
  };
\end{tikzpicture}


\input{Introduction}

\input{Methodologies}

\input{Experiment}

\input{Results}
\input{Discussion}
\input{RelatedWork}
\bibliographystyle{IEEEtran}
\bibliography{ref}


\end{document}

%% file: Introduction.tex
\section{Introduction}

Requirements analysis is a key activity in requirements engineering, where the elicited requirements are reviewed, categorized into different types (e.g., functional and non-functional), dependencies between different requirements are identified, ambiguous requirements are resolved, and then requirements are carefully analyzed to ensure clarity, consistency, and completeness~\cite{1593825}. However, requirements analysis is often laborious and time-consuming. It is particularly challenging, as requirements are often expressed in natural language, which can be susceptible to many forms of \textit{specification defects} and interpretations. For instance, natural language requirements specifications can be inherently \textit{ambiguous} and open to interpretation. These ambiguities can lead to misunderstandings and confusion among stakeholders and may result in incomplete or incorrect requirements. 
Furthermore, it has been shown that the system context and inter-dependencies among multiple requirements can result in conflicts and contradictions among requirements and therefore make requirements analysis more complex~\cite{10.1007/978-3-642-23391-3_3}.
Several studies have attempted to automate specific requirements analysis tasks~\cite{RE14, 10.1145/2507288.2507323,TwinPeaks, 10172729, Jane:Book}. This has resulted in a diverse set of techniques for a variety of tasks such as the classification of requirements into functional and non-functional~\cite{9218141}, identification of sub classes of requirements, traceability and detection of specification issues~\cite{8491122, 10260978}. 
Recent advances in text processing have shown opportunities for leveraging a single method for a variety of inference tasks. In this paper, we investigate the potential of \textit{Natural Language Inference (NLI)}~\cite{bowman-etal-2015-large} as the underlying method to automate requirements analysis tasks. 
NLI, also known as recognizing textual entailment (RTE) is a widely studied method outside of the software engineering domain. 
It determines the relationship between two sentences, known as the \textit{premise} and the \textit{hypothesis}. 
NLI aims to determine whether the premise \textit{entails}, \textit{contradicts}, or is \textit{neutral} to the hypothesis. 
NLI has shown promising results in various tasks, such as \textit{question answering} \cite{demszky2018transforming} and \textit{semantic search}~\cite{stasaski-hearst-2022-semantic}. 
Previous studies have demonstrated the potential of NLI as a universal method for addressing a broad spectrum of NLP tasks, suggesting that various classification challenges can be adeptly reframed into entailment problems, particularly beneficial in data-constrained scenarios such as few-shot and zero-shot learning\cite{wang2021entailment, yin-etal-2019-benchmarking, gera2022zero}. 
This approach relies on changing category labels into detailed sentences, thereby enabling a more nuanced and linguistically coherent assessment of the relationship between text inputs and their corresponding categorizations\cite{sainz-etal-2021-label, wang2021entailment}. 
For the above reasons, NLI can potentially be a promising method for automating a variety of requirements analysis tasks. 

However, despite the impressive results of NLI\cite{jeretic2020natural, wang2021entailment} in producing meaningful and realistic results in prior studies related to genetic text, its practicality for software engineering tasks and more specifically for analyzing software requirements specifications has not been fully explored. 
In this paper we further investigate this hypothesis by designing, developing and evaluating a novel requirements analysis method.
The novel contribution of this paper lies in examining whether the practicality of NLI in other application domains is translatable to the software requirements domain. 
Among existing requirements analysis tasks, we limit the scope of this investigation to three tasks: (1) categorization of requirements to functional and a variety of non-functional requirements, (2) identification of various forms of requirements specification defects and (3) detection of conflicting requirements. 
To further demonstrate the suitability of NLI in the analysis of software requirements specifications, we have conducted a number of benchmark studies using other emerging transformer models and the advancement of large language models (LLMs) such as \textit{ChatGPT}, a \textit{prompt-based model} \cite{10.1145/3551349.3560417, 9218141}, a \textit{conventional BERT model} \cite{9920081, 9218141} along with \textit{conventional statistical classification}~\cite{Jane:AutClassNFR}, in order to assess their efficacy in the realm of software requirements analysis.  This paper investigated the following research questions:

\noindent \textbf{RQ1:} \textit{How effective is NLI in automating the categorization of software requirements?}
Our findings indicate that NLI can accurately automate the process of analyzing and categorizing requirements specifications. NLI can effectively categorize software requirements, achieving an F1 score greater than 83\%, outperforming other methods.
This result highlights that the use of NLI can be a promising approach for automating requirements analysis.

\noindent \textbf{RQ2:} \textit{Can NLI identify and detect various forms of specification defects in software requirements?}
We found that the proposed method can accurately identify specification defects.
Overall, NLI could achieve an F1 score of over 80\% of all defects in requirements specifications, surpassing all other techniques in identifying specification defects.

\noindent \textbf{RQ3:} \textit{How does the proposed approach perform in comparison with other LLM-based methods such as ChatGPT, NoRBERT, Prompt-based classification, as well as a traditional Probabilistic Term Indicator approach?} Our experiments reveal that NLI outperforms all the baseline methods in all three requirements analysis tasks.

\noindent  \textbf{RQ4}: \textit{To what extent can an NLI-based method contribute to the detection of conflicts in requirements?} NLI approach outperformed the baseline techniques, however, the accuracy was lower than other requirements analysis tasks. 


\noindent  \textbf{RQ5}:  \textit{How does the proposed NLI-based approach work in zero-shot learning situations simulating a domain with no previous training data?}
To investigate the practicality of our proposed method for new application domains, scenarios with limited training data, or analysis of a new class of requirements, we evaluate the performance of our method in zero-shot learning scenarios in a cross-project setting.
Our results indicate that our proposed NLI-based method not only outperforms other LLMs in conventional learning settings, but also in zero-shot learning scenarios.

\noindent \textbf{RQ6:} \textit{What lessons can be learned from the application of NLI in software requirements analysis, and how can these lessons inform future research and industry practices?}
We identified a number of practical challenges and limitations of implementing NLI-based approaches. Our experiments reveal that the proposed NLI model highly benefits from label verbalization and knowledge integration that can surpass other techniques in the analyzed requirements engineering tasks.
Additionally, our experimentation demonstrates that we can reformulate inference and classification tasks into entailment problem where NLI is powerful. Additionally, we found that NLI falls short in identifying compositional conflicts among software requirements. Compositional conflicts occur in situations where two requirements are not in conflict, however the interaction of two with a third requirements cause a conflict.


This paper makes the following \textbf{contributions}:

\begin{itemize}
\item An empirically grounded investigation of the NLI method for analyzing software requirements specifications. To the best of our knowledge this is the first time that NLI has been fully investigated in this problem but also in the broader field of software engineering.

\item An automated functional prototype to characterize software requirements specifications, detect specification defects and detect conflicts in requirements.

\item A comprehensive evaluation and comparison of NLI with four additional models for requirements analysis, namely: ChatGPT, NoRBERT, Prompt-based, and Probabilistic Term Indicator. NLI outperforms all the models.


\item Investigating the effect of less studied sub-techniques in the literature such as label verbalization and knowledge integration, and delivering empirical evidence regarding their influence on the model's performance for software requirements classification.
\item Two new datasets containing software requirements specification defects and conflicts in requirements specifications. All datasets are released publicly to the RE community for further advancement of requirements analysis tasks \footnote{Dataset is available at \url{https://zenodo.org/records/11000349}}. 

\end{itemize}

%% file: Methodologies.tex
\begin{table*}[ht]
\caption{Categories of Requirements Defects, Definitions (D), Examples (E), and Color Coding of Root Causes}
\label{tab:requirements-defects}
\centering
\renewcommand{\arraystretch}{1.5} 

\begin{tabular}{|p{1.8cm}|p{15cm}|}
\hline
\textbf{Category} & \textbf{Definition (D) and Example (E)} \\
\hline
\hline
Ambiguous Requirements  & (D) 
The requirements specifications are unclear, imprecise, and open to multiple interpretations. 
(E) Aircraft that are \highlightcyan{non-friendly} \highlightcyan{and} \highlightcyan{have an unknown mission} \highlightcyan{or} \highlightcyan{the potential to enter restricted airspace within 5 minutes} shall raise an alert \\

\hline
Directive Requirements & (D) Direct the developer to look at additional sources beyond the requirement (e.g., referring to a figure or table). (E): ``\textit{The train fault detection module shall follow the \highlightcyan{steps outlined in Figure 1}.}''\\
\hline
Non-Measurable Requirements & (D) The requirements is not measurable, or testable, it contain a qualitative value that is unable to be quantified or measured (e.g., ``some,'' ``large,'' ``long''). (E) ``\textit{The system shall \highlightcyan{scale to  a large group of users} during rush hours.}''\\
\hline
Optional Requirements & (D) Contain terms that may be interpreted in multiple ways and leave the choice of how to implement the requirement to the developer (e.g., ``can,'' ``may,'' ``optionally,'' ``as desired,'' ``at last,'' ``either,'' ``eventually,'' ``if appropriate,'' ``in case of,'' ``if necessary''). (E) ``\textit{The system \highlightcyan{may provide} multi-factor authentication \highlightcyan{whenever necessary}.}''\\
\hline
Uncertain Requirements & (D) Contain a qualitative value that is insufficiently defined (e.g., ``good,'' ``bad,'' ``strong,'' ``easy''), or they may contain under-referenced elements (e.g., ``compliant with standards (which ones?)''). (E) ``\textit{The system must \highlightcyan{deliver fast response times} for user queries, ensuring minimal waiting periods.}''\\
\hline
Non-Atomic Requirements & (D) Describe more than one action to be taken and typically have a conjunction, such as ``and.'' (E) ``\textit{System must add an admin user to its reporting dashboard \highlightcyan{and} allow mobile devices to subscribe to the Reporting flow.}''
\\
\hline
\end{tabular}

\end{table*}
\section{A Natural Language Inference (NLI) based Approach for Requirements Analysis}
\label{sec:NLI}
We set up an experiment pipeline to investigate the effectiveness of NLI~\cite{Guo_Zhang_Liu_2019} for automating requirements analysis tasks. 

\subsection{Task Selection}
Among various requirements analysis tasks, we chose the following tasks due to their prevalence in facilitating the requirements engineering process.

 \subsubsection{Task \#1: Requirements Classification and Characterization} 
    This task involves the analysis of requirements specifications to determine the type and class of the requirements. This process often involves categorizing requirements specifications into functional and non-functional requirements, and additionally identifying sub categories of non-functional requirements such as Security, Performance, Portability, Availability, Fault Tolerance, Legal requirements, Look \& Feel, Maintainability, Operability,  Scalability, Usability ~\cite{Jane:AutClassNFR}.    
    

\subsubsection{Task \#2: Detect Requirements Specification Defects} analysis of requirements specifications to ensure they comply with the best practices of writing good requirements, and making sure specifications are free of errors and ambiguities~\cite{6767215, Fantechi_acontent}.
Several researchers in the RE community have identified different classes of requirements specification defects. Aaron K. Massey et al. \cite{DBLP:conf/re/MasseyRAS14} and Daniel M. Berry et al.~\cite{Berry2004} demonstrated ambiguity as one of the common defects. Alessio Ferrari et a.~\cite{Ferrari} and other researchers ~\cite{Yadla:isse2005,Fantechi_acontent, kamsties2001detecting} have identified various other forms of defects such as \textit{non-measurable}, \textit{non-atomic} and  \textit{uncertain} requirements. Table~\ref{tab:requirements-defects} demonstrates common requirements specification defects discussed in the literature, their definitions, along with sample requirements with such defects. In this paper, we limit the scope of our investigations to these defects.

\subsubsection{Task \#3: Identifying Conflicting Requirements} aims to ensure the consistency and compatibility of software requirements~\cite{mairiza2009managing}. Conflicting requirements is a common problem that occurs when
a software requirement specification is inconsistent with another specification~\cite{10.1007/978-1-4471-0287-8_9}. 
Consistency between requirements specification requires no two or more requirements contradict each other. In this task, the objective is not only to identify conflicting requirements but also to address more complex scenarios. This involves situations where the conflict arises not directly between two requirements but is induced by a third requirement present in the dataset. Table~\ref{tab:conflicting-requirements} lists different types of conflicting requirements with examples.


\begin{table*}[htbp]
\caption{Types of Conflicting Requirements and Color Coding of Root Causes}
\label{tab:conflicting-requirements}
\centering
\begin{tabular}{|m{1.6cm}|m{1.6cm}|m{13cm}|}
\hline
\textbf{Category} & \textbf{Type} & \textbf{Description} \\
\hline
\multirow{5}{1.6cm}{Incompatible Requirements} &  \centering  Two operation frequencies & (D) When the same agent is required to perform the same operation on the same object, but at two different frequencies.(E) ``\textit{R1. The system shall invoice broker accounts \highlightpink{once per year} using CDR.	R2. The system shall invoice broker accounts once per month using CDR.}''  \\
\cline{2-3}
& \centering  Start-forbid & (D)  When the same event causes the same operation to be performed and forbidden. (E)
``\textit{R1. It shall be  \highlightpink{forbidden to delete invoices}.	R2. A broker \highlightpink{shall be able to delete} his own subscribers’ accounts and \highlightpink{associated invoices}}''.
\\
\cline{2-3}
& \centering  Forbid-stop &  When the same operation is stopped under a certain condition event and at the same time, is unconditionally forbidden in another requirement. ``\textit{R1. It \highlightpink{shall be forbidden} for \highlightpink{PBX accounts to have virtual circuits}.	R2. The system shall \highlightpink{deactivate virtual circuits} for \highlightpink{PBX accounts} that do not pay for the service}''.\\
\cline{2-3}
& \centering  Two condition events &  (D) When the same operation is being executed, stopped, or forbidden on two different events. 
(E) ``\textit{R1. The software shall \highlightpink{automatically save the current user session and all unsaved work} \highlightpink{when a loss of network connectivity is detected}. R2. Upon detecting a \highlightpink{loss of network connectivity}, the software shall \highlightpink{immediately terminate the current user session} to maintain data security.}''
\\
\cline{2-3}
& \centering Negation &  (D) When two requirements have a conflicting relationship involving negation. In one requirement, a specific action is mandated, while in the other, the negation or prohibition of that same action is specified. (E)
``\textit{R1. The system \highlightpink{shall store} user information locally. R2. The system \highlightpink{shall not save} user information on a local device temporarily or permanently}''.\\
\hline
\multirow{2}{1.6cm}{\centering Assumption Alteration} & \centering 
 Input-output & (D) When one requirements alternate the assumptions of another requirements. One of the requirements performs its operation on an object (output) that is an input in other requirements. (E) 
``\textit{R1. The buyer shall be able to \highlightpink{view all MLS listings}. R2. The buyer agent shall \highlightpink{make only MLS listing that match the buyers criteria visible to the buyer.}''}
\\
\cline{2-3}
& \centering  Output-output &  (D) This happens if one requirement alters the result (output) or part of another requirement.  (E) 
``\textit{R1. The product shall \highlightpink{support languages of the five target market countries}.	R2. The product shall allow the user to select a chosen language from \highlightpink{a country other than the target market countries}.''}

\\
\hline
\multirow{2}{1.6cm}{\centering Compositional Conflict} & \centering 
 Compositional Conflict &  (D) When two requirements interactions are not causing a conflict, however, when combined with the context provided by a third requirements, a conflict will be formed. (D)
\textit{R1. Once the account is locked, the system \highlightpink{sends an account lock notification email} to the account’s owner. R2. Once an account is locked, the system would also \highlightpink{send a SMS message} to the account’s owner to notify him about the situation owner. R3. If a user has already received a \highlightpink{notification via email}, he will \highlightpink{not receive the same notification via SMS}.} There is a conflict between R1, R2 and R3.
\\
\hline
\end{tabular}

\end{table*}

Below we have described the techniques used in this paper.

\begin{table*}{}
\centering
\caption{Examples of how the data is being prepared for the NLI model. This table shows an example of a sample with [Entail] labels (first row), and an example of a sample with [Contradict] label (second and third rows). To enhance readability, we have opted to employ the original text.}
\label{tab:nli-example}
\begin{tabular}{p{3.2cm}p{1.4cm}p{4cm}p{6.4cm}p{1cm}} 
\hline
\multirow{2}{*}{Requirement (Premise)} & \multirow{2}{*}{Class} & \multirow{2}{*}{Class Description} & \multirow{2}{*}{Final Data} & \multirow{2}{*}{Label} \\ 
& & & & \\
\hline
\hline

\textbf{R.1} The system shall prevent unauthorized connections to it or the remote IoT devices that the system is connected to. & Authorization & This requirement provides authorization to a specific user or group of users for accessing a specific part of the system. &  [\textit{CLS}] The system shall prevent unauthorized connections to it or the remote IoT devices that the system is connected to. [\textit{SEP}] This requirement provides authorization to a specific user or group of users for accessing a specific part of the system. [\textit{EOS}] & Entailment 
\\
\hdashline
\\
\textbf{R.2} The system shall prevent unauthorized connections to it or the remote IoT devices that the system is connected to. & Privacy & This requirement is about keeping user sensitive information private &  [\textit{CLS}] The system shall prevent unauthorized connections to it or the remote IoT devices that the system is connected to. [\textit{SEP}] This requirement is about keeping user specific information private. [\textit{EOS}] & Contradiction 
\\
\hdashline
\\
\textbf{R.3} The system shall be able to call the seller or buyer to schedule an appointment & Functional & This requirement describes the functionality of the system, actions taken by the system or actions the users of the system should be able to take using the system.
 &  [\textit{CLS}] The system shall be able to call the seller or buyer to schedule an appointment. [\textit{SEP}] This requirement describes the functionality of the system, actions taken by the system or actions the users of the system should be able to take using the system. [\textit{EOS}] & Entailment 
\\
\hline
\end{tabular}

\end{table*}

\begin{table*}{}
\caption{Example of how the data is being prepared for the NLI model to detect conflicting requirements.}
\label{tab:nli-Conflict}
\centering
\begin{tabular}{p{4.1cm}p{4cm}p{7cm}p{0.9cm}} 
\hline
\multirow{2}{*}{Requirement (Premise)} & \multirow{2}{*}{Requirement (Hypothesis)} & \multirow{2}{*}{Final Data} & \multirow{2}{*}{Relation} \\ 
& & & \\
\hline
\hline
\vspace{0.1pt} \textbf{R.4} It shall be \highlightcyan{forbidden} for \highlightpink{PBX accounts to have virtual circuits}.	\vspace{0.2pt} & \textbf{R.5} The system shall \highlightcyan{deactivate} \highlightpink{virtual circuits for PBX accounts} that do not pay for the service. &  [\textit{CLS}] It shall be forbidden for PBX accounts to have virtual circuits. [\textit{SEP}] The system shall deactivate virtual circuits for PBX accounts that do not pay for the service. [\textit{EOS}] & \vspace{3pt}Contradict 
\\
\hline
\end{tabular}
\end{table*}

\subsection{The Methodology for applying NLI to Requirements Analysis Tasks}
We exploit the full functionality of NLI by reformulating requirements analysis tasks into entailment or contradiction. For classification we use NLI in a novel way in which we reformulate the underlying requirements classification to an entailment task. 
Our approach was centered around the integration of two key techniques: label verbalization and knowledge integration\cite{sainz-etal-2021-label}.
Specifically, we transformed class names into comprehensive natural language definitions—a process known as label verbalization.
Moreover, by transforming class names into detailed natural language definitions, we made the functionalities more comprehensible, facilitating the integration of domain-specific knowledge into these definitions.
The goal is to determine if the requirement belongs to the given class with the provided class description.
Therefore, the model infers the relationship between the given requirement and the corresponding class description.
To achieve this goal, for each class, we provided a descriptive definition, empowering the model to comprehend the context of the data and leverage the contextual information associated with each label. 
This enhancement aids the model in better understanding and categorizing requirements. 
This converts the first two requirements analysis tasks (NFR-Classification and Defect Detection) into an inference task, for detecting the existence of entailment or contradiction in a given sentence pair. For the third tasks, detection of conflicts in requirements specification, we use NLI in its conventional form, where we infer the conflict relationships between each pair of the requirements. Additionally, we develop an experimentation setup that enables NLI to infer more complex, multi-requirements conflicts. 

\begin{algorithm}
    \caption{Data Preparation Pipeline}
    
    \begin{algorithmic}[1]
        \REQUIRE A dataset of requirements and classes ($D_{input}$)\\
        \hspace*{-2.5em} \textbf{Input:} A dataset of requirements ($R_i$) and the classes ($C$)\\
     \hspace*{-2.15em} \textbf{Output:} \textit{Premise}, \textit{Hypothesis}, and \textit{Label} suitable for the model 
        \STATE $i = 0$
        \WHILE{$i \leq len(D_{input})$}
            \FORALL{$C_j \in C$}
                \STATE let label $L_{i} = 1$, if $R_i$ $\in$ $C_i$, otherwise $L_{i} = 0$
                \STATE Add ($R_i$, $C_i$, $L_{i}$) $\rightarrow$ $D_{output}$, representing \textit{premise}, \textit{hypothesis}, and \textit{label} accordingly
                \ENDFOR   
            \STATE $i = i + 1$
        \ENDWHILE
        \RETURN $D_{output}$
    \end{algorithmic}
    \label{alg:data-preparations}
\end{algorithm}



\subsubsection{Data Preparation}
We proposed a data preparation pipeline for this model, as shown in Algorithm~\ref{alg:data-preparations}. Following the proposed algorithm, the model will create both entailment and contradiction samples that are needed to train the model. This pipeline will transform the sample and its corresponding class into the intended format. 

To create entailment samples, we select requirements and map them to their class, and for creating contradicting samples, for each example, we selected one class that the requirement does not belong to, and added to the dataset.

This pipeline is designed to generate data suitable for NLI model. The dataset comprises samples containing \textit{premises}, \textit{hypotheses}, and a \textit{label} indicating the relationship between the premise and hypothesis (i.e., $1$ for Entailment and $0$ for Contradiction). 
Each requirement is represented once with the corresponding label set to \textit{True}, and $N-1$ times with other labels set to \textit{False}, where $N$ represents the number of classes in the dataset. 
In these experiments, we utilize two labels, diverging from the three-label approach of other NLI models (i.e., Entailment, Contradiction, and Neutral). This choice is driven by our experiments' nature, where each requirement is explicitly assigned to a class or not.


Following the method outlined in Algorithm~\ref{alg:data-preparations}, the dataset has been transformed into a suitable format for model training. 
The \textit{premise} contains the \textit{requirement}, while the \textit{hypothesis} comprises the \textit{descriptive textual representation} of a requirement class (also called label verbalization). 
As a final stage of the data preparation procedure, the data undergoes tokenization to convert into a suitable format for model processing.

For any sample that has followed Equation~\ref{eq:data-tokenized}, the resulting format is $[CLS] \: Requirement \: [SEP] \: Text \: [EOS]$, where $R_{i}$ denotes the tokenized representation of the Requirement (premise), and $T_{i}$ denotes to the tokenized representation of the textual description of the class (hypothesis). 
\begin{equation}
    x_{i} = [CLS] \: R_{i} \: [SEP] \: T_{i} \: [EOS]
    \label{eq:data-tokenized}
\end{equation}

In Equation~\ref{eq:data-tokenized}, \texttt{[CLS]} token denotes the beginning of the input sequence, functioning as an aggregated representation for the entire sequence.
The \texttt{[SEP]} token serves as a separator between different segments of the input, helping the model discern the boundary between the premise and hypothesis. The \texttt{[EOS]} token signals to the model that the end of the sequence has been reached. These tokens collectively define the structure in which the data will be presented to the model.

Table~\ref{tab:nli-example} illustrates an example of how we will make the data suitable for the inference task, using the requirement and class.
As shown, the technique converts the single word class to a descriptive textual description of the class, to provide more information. 
In this table, \textit{Requirement} represents the requirement data, \textit{Class} represents the class corresponding to this data. 
\textit{Class Description}" represents the descriptive text for the corresponding class. Next, the \textit{Requirement} and the \textit{Class Description} will be integrated to form the final data with appropriate format. In the first row, we see an example where the sample belongs to the class, resulting in the final sample being labeled as [Entail]. In contrast, the second row features a class unrelated to the requirement, leading to a [Contradict] label in the final dataset, signifying that the requirement is not classified under this class. The two \textit{Final Sample} and their corresponding \textit{Label}s will be added to the final dataset. Please note that in the Final Samples, there will be a tokenized representation of the text. As each class possesses its own distinct hypothesis, this hypothesis's flexibility empowers the model to grasp the intrinsic information associated with each class more effectively. 

Similarly, Table~\ref{tab:nli-Conflict} demonstrates how the data was prepared for the task 3. This was the conventional use of NLI and unlike task 1 and 2, there was no need for label verbalization.

\subsubsection{NLI Model} Following the proposed pipeline, the processed data will be passed to the model. 
NLI can use variety of architectures such as Long Short-Term Memory (LSTM)~\cite{chen-etal-2017-enhanced}, Recurrent Neural Networks (RNNs)~\cite{Guo_Zhang_Liu_2019}, or Transformer Models such as BERT for the inference task. 
In this paper we use RoBERTa as prior work demonstrated promising performance~\cite{jiang-de-marneffe-2019-evaluating}.
When the processed input is fed into the model, it generates an embedding vector that corresponds to the input provided. Specifically, the model creates a representation for the special \texttt{[cls]} token, which functions as an aggregate representation for the entire sequence. Once these embedding vectors are obtained from the model, they are then directed towards a fully connected neural network classifier. This classifier is responsible for making inferences about entailment or contradiction and ultimately produces the output results.

To this end, we formulate the problem as follows:
Let $x_{i}$ represent the input data which is a representation of the premise (requirement) and hypothesis (class description), as explained in Equation 2 and $y_{i}$ is the entailment label (True / False). The data is fed to the pre-trained model to  generate the vector representation of the input which is passed to the fully connected layers to obtain the output logits.
Sigmoid function is then applied to the logits which maps them to a probability distribution over the two possible labels (entailment or contradiction), $p(\hat{y}_{i,j} \mid x_i, \theta) = \text{sigmoid}(f_{\theta}(x_i))$. The full loss is represented in Equation~\ref{eq:nli-loss}:


\begin{equation}
\begin{aligned}
    Loss =& -\frac{1}{N} \sum_{i=1}^{m}\sum_{j=1}^{n_{i}} \big{[}y_{i,j} \log(p(\hat{y}_{i,j} \mid x_{i,j},\theta))\\
    &+ (1 - y_{i,j}) \log(1 - p(\hat{y}_{i,j} \mid x_{i,j},\theta))
    \big{]}
    \label{eq:nli-loss}
\end{aligned}
\end{equation}

Where $N$ is the total number of the samples in the dataset including $n_{i}$ requirement samples from $m$ classes.

%% file: experiment.tex
\section{Experimentation}
To rigorously evaluate NLI performance in these three tasks, we established a set of benchmark experiments. This section covers the dataset, baseline approaches, and metrics used for evaluation and comparison.

\subsection{Requirements Dataset Selection}
For task 1 (NFR-Classification) we selected the Promise~\cite{promise} dataset, which has been widely used in numerous RE papers~\cite{Jane:AutClassNFR, Cleland-Huang:EarlyAspects, 9218141, kurtanovic2017automatically}. 
We utilized the entire Promise dataset. The class ``Portability'' was excluded from the analysis, following a similar approach as in previous studies~\cite{10.1145/3551349.3560417}, where there was only a single sample available for that class.  The dataset includes 255 functional requirements and 370 non-functional requirements. This dataset, extensively reviewed and used in the RE community, required no additional annotation or peer review.
Table~\ref{tab:promise_dataset} illustrates the classes and the corresponding number of samples for each class. 

\begin{table}[h!]
\centering 
\caption{Number of samples per class in Promise dataset, comprising one class of functional requirements, and 11 classes of non-functional requirements.}
\label{tab:promise_dataset}
\begin{tabular}{|c|c|}
\hline 
\textbf{Class} & \textbf{\# of samples} \\ \hline 
Functional & 255 \\ \hline
Availability & 21 \\ \hline
Fault Tolerance & 10 \\ \hline
Legal & 13 \\ \hline
Look \& Feel & 38 \\ \hline
Maintainability & 17 \\ \hline
Operability & 62 \\ \hline
Performance & 54 \\ \hline
Portability & 1 \\ \hline
Scalability & 21 \\ \hline
Security & 66 \\ \hline
Usability & 67 \\ \hline
\end{tabular}%

\end{table}

For task 2, the requirements specification defect, there is a lack of large publicly available datasets within RE community. Therefore, we established a new dataset of requirements specification defects. To create this dataset, we used several sources. These sources include:
\begin{itemize}
    \item \textbf{Real Software Requirements Documents:} we identified a number of real software systems which their requirements documents were accessible to public, this include systems such as Electronic Health Records System (EHRs) for U.S Department of Health and Human Services.
    \item \textbf{Student Projects:} Requirements were obtained from SysRS documents developed by students taking software engineering courses. These requirements were reviewed and high-quality examples were included in the study.
\end{itemize}

Two subject matter experts (SMEs) from industry with 20+ and 10+ years as requirements analysts were recruited to annotate the requirements through a joint requirements review session where SMEs peer-reviewed and discussed each requirements. The collected requirements undergone through a three-stage review. The requirements were initially analyzed for grammatical and content errors. Then they were labeled by SMEs with various types of specification defects. The labeling sessions was performed in peer-review form, where both SMEs reviewed and discussed the requirements together, and decided about the labels. The labeled requirements were finally peer-reviewed by the authors to verify the annotations and no discrepancies were found. 
As our requirements defects dataset contains defective requirements, these requirements were purposely left containing certain errors. The classes and number of samples of this dataset is shown in Table\ref{tab:defect_dataset}:

\begin{table}[h!]
\centering 
\caption{The classes and samples in the defect dataset.}
\vspace{-3pt}

\label{tab:defect_dataset}
\begin{tabular}{|c|c|}
\hline 
\textbf{Class} & \textbf{\# of samples} \\ \hline 
Ambiguous & 34 \\ \hline
Directive & 4 \\ \hline
Non-Measurable & 18 \\ \hline
Optional & 31 \\ \hline
Uncertain & 16 \\ \hline
Non-Atomic & 28 \\ \hline
\end{tabular}%
\vspace{-3pt}

\end{table}

For Task 3, we constructed a comprehensive dataset by combining seven unique datasets from different sources including:

\begin{itemize}
    \item \textbf{Public Conflict Datasets}: We collelcted three datasets from \cite{10260964}, namely Library, Coffee Machine, and ETCS.
    \item \textbf{Synthesized Conflicting Requirements}: Two SMEs were enlisted to identify and identify or develop a list of conflicting system requirements for a Broker System as well as three projects from the Promise repository (projects \#2, \#6, and \#10). To generate the conflicting requirements the SMEs follow the type of conflicting requirements presented in the Table~\ref{tab:conflict_dataset}.
\end{itemize}
All data was peer-reviewed by the SMEs. The size of different dataset used in Task 3 is illustrated in Table \ref{tab:conflict_dataset}.


\begin{table}[h!]
\centering 
\caption{The number of samples and conflicting pairs in the conflict dataset.}
\vspace{-3pt}
\label{tab:conflict_dataset}
\begin{tabular}{|c|c|c|}
\hline 
\textbf{Dataset} & \textbf{\# of samples} & \textbf{\# of conflicting pairs} \\ \hline 
 Broker & 46 & 13 \\ \hline
 Promise - P2& 65 & 13 \\ \hline
 Promise - P6 &  95 & 14\\ \hline
 Promise - P10 & 69 & 13 \\ \hline
 Coffee Ma. & 21 & 29 \\ \hline
 Library &  109 & 20 \\ \hline
 ETCS & 64 & 34 \\ \hline
\end{tabular}%
\vspace{-3pt}

\end{table}



\subsection{Baseline Approaches for Comparative Study}
This subsection describes the list of baseline approaches we have used for comparison. 

\subsubsection{LLM-based Chatbot}
In this work we aim to examine the performance of an LLM-based chat-bot for the task of requirement analysis. ChatGPT is renowned for its human-like conversational capabilities and has demonstrated considerable potential in the requirement analysis domain~\cite{zhang4450322evaluation}. 
In this study we employed ChatGPT based on GPT-3.5 model.
To assess ChatGPT's performance, we conducted a series of tests using diverse prompts and identified the most effective prompt (the one with the highest F1 score) as detailed below:

``\textit{Given the following definition of a [CLASS\_NAME] requirement:}
\textit{[CLASS\_NAME] : [CLASS\_DESCRIPTION], }

\textit{identify if the following sample belong to class [CLASS\_NAME].}"


\begin{table}[!htb]
\caption{A sample of a ChatGPT prompt. Given the categories and requirements, we executed queries on ChatGPT to determine the samples that belong to each class.}
\vspace{-3pt}
\centering
\small
\begin{tabular}{|p{4cm}|p{4cm}|}
\hline
\textbf{Example1: Prompt}  & \textbf{Example2: Prompt}\\ 
\hline  
\hline 
\textit{Given the following definition of \textbf{Legal} software requirement class:} \newline & \textit{Given the following definition of \textbf{Functional} software requirement class:} \\
\textbf{Legal}:  This requirement describes laws and standards related to the software, its uses, and its users. & \textbf{Functional:} This requirement describes the functionality of the system, actions taken by the system or actions the users of the system should be able to take using the system.  \\
% ...
\textit{identify if the following requirement belong to class scalability with Yes or No.} & \textit{identify if the following requirement belong to class scalability with Yes or No.} \newline  \\
-Requirement: The product shall comply with the estimatics laws relating to recycled parts usage. & -Requirement: The Disputes application shall comply with the corporate standards.
\newline
\newline
\\

\hline 
\textbf{Response} & \textbf{Response} \\
\hline  
\hline  
- Yes & -No
% ...
\\
\hline
\end{tabular}
\label{tab:chatgpt-sample}
\vspace{-3pt}

\end{table}

\subsubsection{NoRBERT, Fine-tuned RoBERTa for Text Classification}
We selected NoRBERT classification model as one of the baselines, as studied by Hey et al.~\cite{9218141} where they introduced a technique for binary classification of FR and NFR, as well as classifying the NFR into the sub-classes. 
However, their approach separates the FR/NFR problem from the sub-classification of NFR, introducing practical challenges. 
This separation in their methodology leads to unrealistic results, as the accuracy of sub-classifying NFRs is contingent upon the initial distinction between FRs and NFRs.
In contrast, our setting involves directly categorizing requirements into functional and sub-classes of non-functional, mirroring a more realistic scenario in requirements engineering tasks.

The baseline used in this paper is trained to perform multiple binary classifications where a separate model is trained for each class. Formally, let $x_{i,j}$ represent the $j^{th}$ sample from the class $i$ and $y_{i,j}$, its corresponding label. We denote $f_{\theta}(x_{i,j})$ as the output logits, which are raw predictions produced by the fine-tuned RoBERTa classification model with learnable parameters $\theta_i$. The Sigmoid function is then applied to these logits to obtain class probabilities.
Thus, the loss for each model $Loss_{i}$, will be defined as shown in \eqref{bert-loss},
\begin{equation}
\begin{aligned}
    Loss_{i} =& -\frac{1}{n_{i}} \sum_{j}^{n_{i}} \big{[}y_{i,j} \log(p(y_{i,j} \mid x_{i,j}, \theta_{i})) + \\
    & (1 - y_{i,j}) \log(1 - p(y_{i,j} \mid x_{i,j}, \theta_{i}))\big{]}
    \label{bert-loss}
\end{aligned}
\end{equation}

where $n_{i}$ is the number of training samples belonging to the class $i$.

\subsubsection{Prompt-based Classification}

Text classification using prompts~\cite{wang-etal-2023-prompt} is a technique that allows the model to bridge the gap between the pretraining and the downstream task by its flexibility, as it can fine-tune the model's behavior by designing prompts tailored to the task at hand. 
For instance, for any requirement and the corresponding class, the sample with the added prompt will be as follows: ``\{Requirement\}. This requirement is \texttt{[MASK]}."
In this method, the first part of the input contains the requirement, and the model is trained to predict the masked class name within the given prompt. Recently,
~\cite{10.1145/3551349.3560417} 
proposed an alternative to predicting the masked class name in the prompt. They showed that a binary classification model trained to predict whether the given prompt matches the class or not, would outperform the mask prediction model. This technique serves as another baseline in this paper. 
Thus, the processed input  follows this format: ``\{Requirement\}. This requirement is \texttt{[Class\_Name]}." Here, the input is a requirement augmented with the class name, referred to as a prompt in this technique, and the output is a binary indicator reflecting whether the sample falls into the category specified in the prompt or not.
Similar to the previous approach, this baseline is a fine-tuned RoBERT-based model, aiming to minimize the loss between the predicted label for the given prompt and the true labels of the sample. The model is validated on a separate validation set, and tested on the unseen test set. Given the tokenized requirement sequence $x_{i,j}$ and its corresponding ground truth label $y_{i,j}$, $p_{i,j} $ represent the updated prompt consisting of $x_{i,j}$ augmented with the description of the class $y_{i,j}$. We denote $f_{\theta}(p_{i,j})$ as the output logits of the fine-tuned prompt-based model with learnable parameters $\theta$. The Sigmoid function is then applied to these logits to obtain class probabilities as $p(\hat{y}_{i,j} \mid p_i, \theta) = \text{sigmoid}(f_{\theta}(p_i))$. 
Thus, the loss will be defined as shown in \eqref{eq:prompt-loss},
\begin{equation}
\begin{aligned}
    Loss =& -\frac{1}{N} \sum_{i=1}^{m}\sum_{j=1}^{n_{i}} \big{[}y_{i,j} \log(p(\hat{y}_{i,j} \mid p_{i,j},\theta))\\
    &+ (1 - y_{i,j}) \log(1 - p(\hat{y}_{i,j} \mid p_{i,j},\theta))
    \big{]}
    \label{eq:prompt-loss}
\end{aligned}
\end{equation}
where $m$ is the number of classes, $n_{i}$ is the number of requirements per class and $N$ is the total number samples in the dataset.

\subsubsection{Probabilistic Indicator Terms-Based Approach}
The last baseline method is a seminal classification technique developed by Cleland-Huang et. al.~\cite{Cleland-Huang:EarlyAspects, DBLP:journals/tse/Cleland-HuangCC03} specifically for the task of requirements analysis~\cite{Jane:AutClassNFR, Cleland-Hang2007, RE14, Jane:Book} and classification~\cite{TSE2016, ICSE2012, FSE2014}. This technique utilizes a probabilistic algorithm created for classification. As a terms-based approach, it is best suited towards data where each category is more likely to use a set of specific terms. It consists of preprocessing, followed by a two-phase classification approach. 

The pre-processing step consists of removing ``non-textual characters'' such as punctuation and numbers, removing duplicate whitespace, converting all characters to lowercase, tokenizing, stopword removal, and stemming the remaining words using the Porter algorithm \cite{porter1980algorithm}.

In the first step of the classification process, a set of \textit{indicator terms} are identified for each label category. Indicator terms are tokens assigned a probabilistic weight based their occurrence within each category's requirements. More significant indicator terms (with a higher weight) represent terms that are \textit{more likely} to occur for a specific category, while less significant ones are \textit{less likely}.
In the second step, requirements are classified by using the weighted indicator terms obtained in the first phase to determine the probability that each belongs to a specific category.


\begin{table*}[!h]
\centering
\caption{The results of various models on the promise dataset. All numbers are in percentage (\%).}
\vspace{-3pt}
\label{tab:results}
\resizebox{\linewidth}{!}{%
\begin{tabular}{|l|ccc|ccc|ccc|ccc|ccc|}
\hline
\multirow{2}{*}{Class} & \multicolumn{3}{c|}{\textbf{ChatGPT}} & \multicolumn{3}{c|}{\textbf{NoRBERT}} & \multicolumn{3}{c|}{\textbf{Prompt-Based}} & \multicolumn{3}{c|}{\textbf{NLI}} & \multicolumn{3}{c|}{\textbf{Indicator Term Freq.}} \\ 
 & Prec. & Rec. & F1 & Prec. & Rec. & F1 & Prec. & Rec. & F1 & Prec. & Rec. & F1 & Prec. & Rec. & F1 \\ \hline
Scalability & 33 & 67 & 44 & 22
& 33
& 27
& 72 & 67 & 69 & 89 & 78 & \cellcolor{lightblue}82 & 33 & 76 & 46 \\ 
Legal & 67 & 100 & 80 & 89
& 83
& 82
& 48 & 57 & 52 & 89 & 89 & \cellcolor{lightblue}89 & 20 & 40 & 27 \\ 
Performance & 55 & 86 & 67 & 93
& 89
& 90
& 89 & 89 & 88 & 93 & 89 & \cellcolor{lightblue}91 & 44 & 47 & 46 \\ 
Security & 75 & 86 & 80 & 69
& 75
& 72
& 78 & 90 & 83 & 92 & 84 & \cellcolor{lightblue}86 & 74 & 69 & 44 \\
Operability & 45 & 90 & 60 & 75
& 90
& 82
& 82 & 82 & 82 & 94 & 78 & \cellcolor{lightblue}86 & 29 & 29 & 29 \\
Usability & 71 & 83 & 77 & 81
& 90
& 85
& 74 & 85 & 79 & 90 & 86 & \cellcolor{lightblue}88 & 32 & 59 & 42 \\
Maintainability & 50 & 100 & 67 & 67
& 100
& 78
& 67 & 50 & 57 & 93 & 83 & \cellcolor{lightblue}85 & 14 & 53 & 23 \\
Functional & 64 & 74 & 68 & 94
& 88
& \cellcolor{lightblue}91
& 87 & 75 & 80 & 93 & 59 & 71 & 52 & 67 & 59 \\
Fault Tolerant & 50 & 100 & 67 & 33
& 33
& 33
& 61 & 72 & 66 & 93 & 83 & \cellcolor{lightblue}85 & 32 & 57 & 41 \\
Look \& Feel & 80 & 100 & \cellcolor{lightblue}89 & 89
& 83
& 86
& 81 & 72 & 76 & 93 & 70 & 80 & 25 & 62 & 36 \\
Availability & 50 & 100 & 67 & 100& 53& 63& 82 & 89 & 85 & 92 & 69 & \cellcolor{lightblue}78 & 16 & 85 & 27 \\
\hline
\end{tabular}
}\vspace{-3pt}

\end{table*}

\begin{table*}[!t]  
\renewcommand{\arraystretch}{1.3}  
\caption{Detecting Requirements Specification Defects - Performance Comparison of Models. All numbers are in percentage\%}
\label{tab:results-standard-def}
\vspace{-3pt}
\centering
\begin{tabularx}{\textwidth}{|l|*{18}{>{\centering\arraybackslash}X|}}
\hline
\rowcolor{headercolor}
Model & \multicolumn{3}{c|}{Ambiguous} & \multicolumn{3}{c|}{Directive} & \multicolumn{3}{c|}{Non-Meas.} & \multicolumn{3}{c|}{Optional} & \multicolumn{3}{c|}{Uncertain} & \multicolumn{3}{c|}{Non-Atomic} \\
\cline{2-19}
 & \textbf{Prec.} & \textbf{Rec.} & \textbf{F1} & \textbf{Prec.} & \textbf{Rec.} & \textbf{F1} & \textbf{Prec.} & \textbf{Rec.} & \textbf{F1} & \textbf{Prec.} & \textbf{Rec.} & \textbf{F1} & \textbf{Prec.} & \textbf{Rec.} & \textbf{F1} & \textbf{Prec.} & \textbf{Rec.} & \textbf{F1} \\
\hline
ChatGPT & 44 & 82 & 57 & 50 & 33 & 40 & 61 & 61 & 61 & 56 & 81 & 66 & 48 & 94 & 64 & 46 & 57 & 51 \\
NoRBERT & 17 & \cellcolor{lightblue}100 & 29 & 67 & 67 & 67 & 53 & 50 & 51 & \cellcolor{lightblue}75 & \cellcolor{lightblue}100 & \cellcolor{lightblue}86 & 75 & 100 & 86 & 51 & 71 & 60 \\
Prompt-based & 58 & 82 & 68 & 75 & 100 & 86 & \cellcolor{lightblue}63 & 67 & 65 & 72 & 84 & 78 & 73 & 100 & 84 & 50 & 64 & 56 \\
NLI & \cellcolor{lightblue}74 & 78 & \cellcolor{lightblue}76 & \cellcolor{lightblue}100 & \cellcolor{lightblue}100 & \cellcolor{lightblue}100 & 50 & \cellcolor{lightblue}100 & \cellcolor{lightblue}67 & 67 & 100 & 80 & \cellcolor{lightblue}100 & \cellcolor{lightblue}100 & \cellcolor{lightblue}100 & \cellcolor{lightblue}53 & \cellcolor{lightblue}82 & \cellcolor{lightblue}65 \\
\hline
\end{tabularx}
\vspace{-3pt}
\end{table*}


\subsection{Setup and Parameters}
For experiments, we utilized Python 3.9, PyTorch 1.13.1, and an NVIDIA T4 with 16GB memory and 4 vCPUs.
For the NoRBERT model, prompt-based model, and NLI tasks, we used RoBERTa model~\cite{liu2019roberta}.  
We trained on the training set, monitored the best model with the validation set, then evaluated it on the unseen test set.
We partitioned the dataset into training, validation, and test sets in an 80:10:10 ratio. Validation tracks and selects the best model, based on the lowest loss across epochs.
Upon completing the training process, we finally assessed the model's performance on the test set. 
For these three techniques we trained the model with 50 epochs, and we used the AdamW optimizer, with testing different learning rates \{$1e-2, 1e-3, 1e-4, 1e-5, 1e-6, 1e-7$\}, where $learning\_rate = 1e-5$ consistently achieved the best results.
For the probability-based approach, we tested the following values as the classification thresholds: [1, 3, 5, 10, 15, 20, 25, 30, 35, 40]. For indicator term thresholds, we tested various values, and selected the following values: [0.001, 0.0001, 0.00001]. During the experiment, we followed 10-fold cross validation.
For ChatGPT, we conducted each experiment in a new chat window to avoid any potential influence from prior experiment queries.  We queried the model with one sample per query, and asked ChatGPT to generate the classification results using the described query.
In the experiments, for datasets exhibiting imbalance, we utilized oversampling and undersampling techniques address the challenge of imbalance, and the test set remained unaltered.
During testing, all models were tested using a binary format for each individual class.

\subsection{Metrics}
The metrics used in this paper to show the effectiveness of the proposed approach are: \textit{Accuracy, Precision, Recall,} and \textit{F1}.
We defined them using True Positives (TP), False Positives (FP), True Negatives (TN), and False Negatives (FN). In the definitions provided, "positive" denotes entailment, while "negative" indicates contradiction.
\begin{itemize}
    \item TP: Number of samples correctly classified as positive 
    \item FP: Number of samples incorrectly classified as positive
    \item TN: Number of samples correctly classified as negative
    \item TN: Number of samples incorrectly classified as negative
    
\end{itemize}

Then, the four measures can be explained as:
\begin{itemize}
    \item \textit{$Accuracy = (TP + TN)/(TP + TN + FP + FN)$}
    \item \textit{$Precision = TP / (TP + FP)$}
    \item \textit{$Recall = TP / (TP + FN)$}
    \item \textit{$F1 = 2*Precision*Recall/(Precision + Recall)$}
\end{itemize}

To report the final scores, we utilized a weighted average for each score, assigning a weight equivalent to the sample size associated with each label. 
This approach ensures that scores are proportionally represented based on the number of samples belonging to each category.

%% file: Results.tex
\section{Analysis and Results}
We measure the performance of NLI and the baseline methods for three requirements analysis tasks: (1) requirements classification, (2) detection of requirements specification defects and (3) discovering conflicting requirements. 
In this experiment, we have presented the scores for each individual class to facilitate comparison of the performance of various models within each specific category.
Each model has been tested three times, and the average scores are reported.

\begin{framed}
\noindent
\textbf{RQ \#1:} The NLI method accurately classifies requirements, achieving an overall F-1 is 83\%.
\end{framed}


Table~\ref{tab:results} summarizes various accuracy metrics for Task 1. 
These metrics are reported per-class and per-benchmark method. 
As the table indicates, NLI surpasses other methods across all categories in terms of F1 score, with the average of 83\% across the entire dataset, with the highest of 91\% in class Performance, except two classes, Functional and Look \& Feel categories which NoRBERT and ChatGPT secured the highest F1 scores, outpacing NLI by margins of 20\% and 9\%, respectively.
It worth to mention that, prior work on requirements classification have focused either on classifying requirements to FR and NFR or classifying NFRs into sub-classes. This assumption while providing a higher accuracy, is rather unrealistic, since in real-life situations we may not have all NFR previously separated from functional one. In our experiment, we tackled a more complex task, in which all the classes including FR were incorporated in the experiments.

Table~\ref{tab:results-standard-def} represents the results of models on the defect dataset for Task 2. The term frequency model was not tested on the defect dataset because many of the defects originate from a requirement's grammatical structure, and not the specific terms contained within the requirement. 
NLI could outperform other techniques in Task 2 across all classes in terms of F1 score with a big margin, except for Optional class, in which NoBERT model achieved the highest F1 score with a 6\% margin compared to NLI.
NLI could achieve 100\% F1 score on two of the classes, Uncertain and Directive, with a 16\% and 14\% difference from the next best performing model.

 \begin{framed}
\noindent
\textbf{RQ \#2:} The NLI method accurately performs the task of detecting requirements defects. The overall F-measure is above 80\% across the dataset.
\end{framed}

Furthermore, the comparison of the NLI with other approaches demonstrates that the NLI method outperforms all other baseline methods for task 1 and task 2.
This is notable, especially given its significantly smaller size relative to the ChatGPT model and the fact that it shares the same pretrained language model as both NoRBERT and prompt-based techniques.

 \begin{framed}
\noindent
\textbf{RQ \#3:} The NLI method demonstrated superior performance compared to other LLM-based methods
such as ChatGPT, NoRBERT, Prompt-based model, as well as traditional Probabilistic Term Indicator approach.
\end{framed}

For task 3, we only compared NLI and ChatGPT, as the nature of task and data was not compatible with the remaining models. Table~\ref{tab:Conflictperformance_comparison} summarizes the finding. The overall accuracy is lower than the other tasks. Overall our data included many non-trivial examples of conflicts. For instance, we only had one example of simple negation, the remaining conflicts required a deep understanding of each requirements and their relationships with other requirements. 

\begin{framed}
\noindent
\textbf{RQ \#4:} The NLI method outperformed ChatGPT for the task of discovering conflicts among requirements, however, the overall accuracy remained suboptimal.
\end{framed}


\subsection{Zero Shot Experiments}

 \begin{table*}[ht]
\centering
\caption{Discovering Conflicts in Requirements: Comparing models across different datasets in Zero Shot Setting. All numbers are in percentage \%}
\label{tab:Conflictperformance_comparison}
\vspace{-3pt}
\renewcommand{\arraystretch}{1.2} 
\setlength{\tabcolsep}{4pt} 
\begin{tabular}{|l|*{21}{c|}}
\hline
\multirow{2}{*}{Method} & \multicolumn{3}{c|}{Brkr.} & \multicolumn{3}{c|}{Lib.} & \multicolumn{3}{c|}{Cof. Mach.} & \multicolumn{3}{c|}{ETCS} & \multicolumn{3}{c|}{Prom.-P2} & \multicolumn{3}{c|}{Prom.-P6} & \multicolumn{3}{c|}{Prom.-P10} \\ \cline{2-22}
 & Prec. & Rec. & F1 & Prec. & Rec. & F1 & Prec. & Rec. & F1 & Prec. & Rec. & F1 & Prec. & Rec. & F1 & Prec. & Rec. & F1 & Prec. & Rec. & F1 \\
\hline
ChatGPT & 42 & 23 & 30 & 20 & 10 & 13 & 66 & 13 & 22 & 33 & 6 & 10 & 12 & 7 & 9 & 20 & 8 & 11 & 14 & 7 & 10 \\
NLI & 67 & 46 & \cellcolor{lightblue}55 & 19 & 80 & \cellcolor{lightblue}31 & 57 & 45 & \cellcolor{lightblue}50 & 27 & 50 & \cellcolor{lightblue}35 & 44 & 31 & \cellcolor{lightblue}36 & 50 & 23 & \cellcolor{lightblue}31 & 50 & 14 & \cellcolor{lightblue}22 \\
\hline
\end{tabular}
\end{table*}

We examine the efficacy of NLI in zero-shot setting contexts, in Task 3, aiming to explore its potential for automating requirements analysis. This investigation targets new sub-types across distinct application domains, particularly those constrained by limited training data.
Our methodology follows with the proposed settings for testing NLI in zero shot scenarios in \cite{wei2021finetuned, plaza2022natural}. We adapted the model by fine-tuning it on all pairs of requirements from N-1 projects, where N is the the total number of projects. 
Subsequently, the model is evaluated on the requirement pairs from remaining project.  This setting ensures that the NLI model is assessed against a new set of hypotheses and has not been exposed to the requirements or the hypotheses from the excluded project.

For comparison,  the model is envisioned to analyze pairs of requirements. The model receives two requirements as inputs, and generate the results. 
Among the available techniques, besides NLI, ChatGPT can be readily employed to identify conflicting pairs. 
We have also experimented with other methods, which involve combining and presenting the two requirements as a single input to their respective models. However, due to their significantly low accuracy, often zero, we have decided to exclude those results from our analysis.

As shown in Table \ref{tab:Conflictperformance_comparison}, NLI consistently achieved higher F1 scores than ChatGPT in all seven projects, even though its overall performance was not particularly outstanding. 



 \begin{framed}
\noindent
\textbf{RQ \#5:} The NLI method demonstrated superior performance compared to ChatGPT in a zero-shot learning scenario. This finding is particularly noteworthy as it not only showcases NLI's capacity to mitigate the challenges of traditional supervised learning, such as the requirement for extensive labeled datasets and the difficulty in generalizing to novel classes.
\end{framed}
\subsection{Lessons Learned}
This paper highlights several lessons learned discovered through experimental use of NLI for three requirements analysis tasks. 

\textbf{Reformulating Requirements Analysis Tasks through Label Verbalization:} 
Label verbalization, refers to the process of transforming the class labels into a form that can be understood and processed by the model. This often involves converting categorical class labels into rich, descriptive language that the model can use to relate to its learned representations \cite{sainz-etal-2021-label}. Our experiments reveal that the proposed NLI model highly benefits from label verbalization and knowledge integration that can surpass other techniques in the analyzed requirements engineering tasks. Moreover, NLI can perform well in the settings with limited data \cite{wang2021entailment}.
In particular we exploited label verbalization along with reformulation of classification tasks into entailment or contradiction to fully benefit from NLI. 
The results indicate such approach to be promising.

\textbf{Multi-Requirements Inter-dependencies are Hard to Capture:} NLI could not detect composite conflicts causes due inter-dependencies among more than two requirements. This was expected as NLI only investigated the entailment or contradiction between premise and hypothesis, where each were a single specification. On the other hand, ChatGPT was able to find some of the composite conflicts. This was because, GPT model had access to all requirements together and was able to explore inter-dependencies. For NLI to achieve similar results, it may be possible to combine multiple requirements as premise to explore the interaction between more than two requirements. However, we did not explore such setting.

\textbf{Prompts and Verbalization are Important:}
Prompt design is a crucial task in ChatGPT~\cite{10260978}. It is essential to recognize that one general rule does not universally apply to all datasets, and each dataset may perform better with a specific prompt tailored to its characteristics. Similarly in NLI, verbalization is important. The effectiveness of the verbalization depends on how well the descriptive terms capture the essential characteristics of the class and how well they align with the features learned by the model. 

 \begin{framed}
\noindent
\textbf{RQ \#6:} NLI model highly benefits from label verbalization and knowledge integration. Tasks in requirements analysis that require inference and classification can be reformulated into an entailment problem, where NLI is highly effective. 
NLI falls short in identifying compositional conflicts among software requirements, which ChatGPT partially succeeds. 
\end{framed}

%% file: Discussion.tex
\section{Threats and Limitations}
In this section, we address potential threats to the validity of our research and acknowledge its limitations.

\textit{Construct Validity Threats:} There is a possibility that the prompts and verbalized labels utilized in our experiments may not accurately encapsulate the constructs they are intended to measure. This raises concerns about the precision with which these tools assess the capabilities of LLMs in requirements engineering. To mitigate this threat we tried to utilize a generic class description for verbalization and heuristically discover the best prompt template for ChatGPT. It's important to acknowledge that, despite our comprehensive experiments, the range of prompts and verbalized labels examined when working with LLMs may not necessarily represent their optimal performance. Therefore, caution is advised when generalizing these techniques to other tasks and domains.

\textit{Internal Validity Threats:} Our investigation might be subject to the effects of data memorization—a phenomenon inherent to LLMs during the training phase—which could influence the causal relationships under scrutiny. The performance metrics derived from our experiments might be artifacts of such confounding factors rather than genuine indicators of model efficacy.

\textit{External Validity Threats:} the generalizability of findings is limited to the datasets included in this study. The specific requirements datasets, their complexities, interconnections and the application domains may not capture the broader spectrum of challenges that may exists in requirements analysis tasks.

\section{Discussion \& Conclusion}
\label{sec:disc}
\vspace{-3pt}
In this work, we investigate the potentials of NLI across three different tasks within requirements engineering: categorized requirements into functional and various non-functional, identifying different forms of requirements specification defects, and detection of conflicting requirements, and compared the performance of NLI against various benchmarks. Besides, we presented two new datasets: one for software requirements specification defects and another for conflicts in requirements specifications.
For Task \#1, unlike many past studies which first classified FRs and NFRs, then classify NFRs into sub-classes--a process that resulted inflated accuracy metrics, our approach aims to directly classify requirements into FRs and sub-classes of NFRs. This approach mirrors more realistic scenarios where FRs and NFRs often intermingle, thereby providing a more accurate assessment of different methods in our study. For Task 1 and 2, we reformulated the classification task into an entailment problem. This transformation allows to employ label verbalization, allowing the model to leverage the knowledge embedded within class descriptions. 
Our experiments shows the effect of verbalization on the models performance as we compare the results with the NoRBERT and prompt-based model. Additionally, the results highlight how reformulating classification to entailment can enhance the analysis of requirements. Despite ChatGPT's impressive capabilities in general NLP tasks, we have observed that it is not the optimal approach for software requirements analysis compared to other techniques. Especially, our experiments reveal that smaller-sized NLI models can exceed ChatGPT's performance in all three evaluated tasks, consistent with the findings of previous studies conducted in various domains\cite{pahwa-pahwa-2023-bphigh, wang-etal-2024-rethinking}.

The superiority of prompt-based approaches over NoRBERT is apparent because they offer small, yet effective, contextual information relevant to the task at hand, as they contains the name of the class in their prompt added to the samples.
Integrating domain-specific knowledge significantly enhances the NLI method compared to the prompt-based approach. The NoRBERT model, the prompt-based approach, and NLI all build on the same pre-trained model. However, the integration allows for the inclusion of comprehensive information within samples, overcoming the constraints of relying on single-word prompts for predictions.
This enables the model to leverage descriptive information within the hypothesis for the inference task, resulting in more accurate identification of the relationship between the hypothesis and premise in the inference task. 

In Task \#3, our experiments demonstrate that the NLI method surpasses ChatGPT in identifying conflicting requirements in a given set of requirements. Although NLI shows superior performance, it struggles to recognize compositional conflicts—those arising from interactions with other requirements in the dataset. This limitation stems from NLI's focus on analyzing pairs of sentences and determining how they are related. A potential area for future research is exploring how NLI might be adapted to effectively incorporate additional contextual information to detect multi-dimensional conflicts among three or more requirements.


\vspace{-2pt}
\section{Environmental Consideration, Resource Consumption and Efficiency}
\vspace{-2pt}
The exponential expansion of machine learning techniques has sparked significant apprehensions regarding their environmental impact and carbon footprint. 
Prioritizing energy-efficient models and optimizing algorithms can noticeably mitigate carbon emissions~\cite{9810097}. Further research is required to better understand the technical advantage of such automated requirements analysis techniques in relation to the resource consumption and impact on environments. 
Particularly, our experiments reveal that despite NLI's significantly smaller size compared to the ChatGPT model, and its use of the same foundational language model as NoRBERT and prompt-based methods, NLI demonstrates noteworthy capabilities.
The integration of robust sustainable software engineering practices and an ethos of knowledge-sharing within the software engineering and artificial intelligence community are essential in our quest to minimize the ecological ramifications of artificial intelligence technologies. 
With collaborative endeavors actively directed towards addressing this issue, we can forge a path towards a more sustainable and responsible future for software engineering.

%% file: RelatedWork.tex
\section{Related Work}
\vspace{-2pt}
The classification of software requirements is a longstanding, yet well-studied problem~\cite{Mahmoud:2016, Jane:AutClassNFR}. 
Tong Li et al.~\cite{LI2020110566} proposed an ontology-based classification method for security requirements. Built on top of the defined formal security ontology, they defined a set of keywords for each class, and proposed a keyword-based classification method. They proposed a methodology for comparing and contrasting requirements modeling languages based on ontological principles. The methodology uses Term Frequency-Inverse Document Frequency (TF-IDF) to calculate and rank the frequency of words in each document to map natural language descriptions of ontology concepts to requirements modeling languages, and was compared with Naïve Bayes (NB), Decision Tree, and Linear Regression.
Feng-Lin Li et al.~\cite{6912271} proposed quality-based language modeling for non-functional requirements. 
Cleland-Huang et al.~\cite{Cleland-Hang2007} proposed a probabilistic-based retrieval method for non-functional requirements classification based on a set of terms related to each class. 
Abad et al.~\cite{Abad:2017} conducted a study that contributed a preprocessing that can further improve the results of the functional and non-functional requirements classification. They compared different classification methods, including Binarized Naïve Bayes, a frequency based-algorithm, and two unsupervised techniques including k-means and hierarchical clustering. 
With the progress of LLMs, previous studies have explored the potentials of transferring these models to requirements classification tasks. Hey et al.~\cite{9218141} presented NoRBERT to identify the capabilities of fine-tuned RoBERTa model in requirements classification. 
Knauss et al. proposed a Bayesian classifiers for identifying security requirements for domain-independent projects \cite{Knauss:2011}, and extracted a set of keywords and utilized classifier to heuristically categorize requirement statement into \textit{security-relevant} and \textit{less-security-relevant} classes. Munaiah et al. created a domain-independent classifier that determines if requirements are security-related or not \cite{Munaiah:2017}. They used TF-IDF to collect unique words to use for class identification. 
They used the One-Class SVM model on the CWE and ExCWE datasets for training.
They found that this classifier outperformed the Naïve Bayes classifiers from previous literature. They also found that their dataset yielded better performance than using the dataset from Knauss et al.~\cite{Knauss:2011}. 
In the literature there are a number of heuristic and pattern based methods for detecting requirements specification defects~\cite{6767215, Fantechi_acontent}. Additionally, researchers have defined a set of rules~\cite{6767215} to detect requirements that contain i) simple sentences, ii) declarative sentences, and iii) active sentences and separate them from non-atomic ones. Other researchers have developed \textit{analytical keywords and information retrieval techniques} to detect uncertain requirements defects~\cite{Fabbrini,610237,992662}. 

NLI has been widely studied in various domains and applications. Goldzycher et al.\cite{goldzycher-etal-2023-evaluating} studied potentials of NLI in hate speech detection and Yant et al.\cite{yang2019fake} studied fake news detection as an inference task. 
Similarly, NLI has been applied to the field of misinformation detection\cite{hossain2020covidlies, arana-catania-etal-2022-natural}, clinical domain\cite{romanov-shivade-2018-lessons, yang2022large}, text summarization and question answering \cite{demszky2018transforming, mishra2021looking}. Fantechi et al. \cite{10260964} have studied the performance of ChatGPT in identifying the inconsistencies in software requirements. They employed ChatGPT to generate conflicting requirements, thus ChatGPT had prior exposure to the conflicting samples. In our study, the conflicting samples have not been previously exposed to ChatGPT, representing a more realistic scenario for identifying the performance of ChatGPT.  Zhang et. al. \cite{zhang4450322evaluation} previously investigated the performance of ChatGPT in requirement information retrieval tasks. They studied the performance of ChatGPT in zero shot shot setting for the four NFR classes of \textit{Security, Performance, Usability, Operational}, with the following prompt: \textit{Tag one quality label from (Usability, Security, Operational, Performance) for the following non-functional requirement statement}. Thus, exploring the effectiveness of the model in identifying other classes of NFRs as well the FRs remains unexplored, which we address in this work.